\documentclass[aps,prl,twocolumn,showpacs]{revtex4-1}
\usepackage{hyperref}  % needed for hyperlinks (doi and equation numbers)
\hypersetup{
pdfauthor={Jerome C. Sanders, Otim Odong, Juha Javanainen, Matt Mackie},
pdftitle={Bound states of two bosons in an optical lattice near an
association resonance},
colorlinks=true,
urlcolor=blue,
linkcolor=blue,
bookmarks=true,
pdftoolbar=true,
filecolor=red,
citecolor=blue
}
\usepackage{graphicx}  % needed for figures
\usepackage{dcolumn}   % needed for some tables
\usepackage{bm}        % for math
\usepackage{amssymb}   % for math
\usepackage{verbatim}  % for comment
\usepackage{epstopdf}
\usepackage{color}
\newcommand{\ket}[1]{\ensuremath{\left| #1 \right\rangle}}

\newcommand{\abs}[1]{\ensuremath{\left| #1 \right|}}

\newcommand{\beq}{\begin{equation}}
\newcommand{\eeq}{\end{equation}}
\newcommand{\bea}{\begin{eqnarray}}
\newcommand{\eea}{\end{eqnarray}}

\newcommand{\eq}[1]{{(\ref{#1})}}
\newcommand{\commentout}[1]{{}}

\newcommand{\half}{{\hbox{$\frac{1}{2}$}}}
\newcommand{\Kappa}{{\cal K}}
\newcommand{\K}{{\Xi}}

\definecolor{red}{rgb}{1,0,0}

\begin{document}
\title{Bound states of two bosons in an optical lattice near an association resonance}
\author{Jerome C. Sanders}
\author{Otim Odong}
\author{Juha Javanainen}
\affiliation{Department of Physics, University of Connecticut, Storrs, Connecticut 06269-3046, USA}
\author{Matt Mackie}
\affiliation{Department of Physics, Temple University, Philadelphia, Pennsylvania 19122, USA}

\begin{abstract}
We model two bosons in an optical lattice near a Feshbach or photoassociation resonance, focusing on the Bose-Hubbard model in one dimension. Whereas the usual atoms-only theory with a tunable scattering length yields one bound state for a molecular dimer for either an attractive or repulsive atom-atom interaction, for a sufficiently small direct background interaction between the atoms a two-channel atom-molecule theory may give two bound states that represent attractively and repulsively bound dimers occurring simultaneously. Such unusual molecular physics may be observable for an atom-molecule coupling strength comparable to the width of the dissociation continuum of the lattice dimer, which is the case, for instance, with narrow Feshbach resonances in Na, $^{87}$Rb, and $^{133}$Cs or low-intensity photoassociation in $^{174}$Yb.
\end{abstract}

% 03.75.Lm (03.75.-b is Matter waves) Tunneling, Josephson effect, Bose-Einstein condensates in periodic potentials, solitons, vortices, and topological excitations
% 37.10.Jk Atoms in optical lattices
% 05.30.Jp Boson systems
% 05.50.+q Lattice theory and statistics

\maketitle

An optical lattice~\cite{MOR06} modifies the motion of atoms profoundly compared to free space. Moreover, the site-to-site hopping of the atoms can be controlled by varying the intensity of the lattice laser, which enables phenomena ranging from Mott insulators~\cite{Jaksch81} to atomtronics~\cite{PEP09}. The interactions between the atoms can also be tuned by utilizing Feshbach~\cite{FAT08, POL09} and photoassociation~\cite{THE04,YAM10} resonances. In effect, both the mass of the atoms and the atom-atom interactions can be controlled experimentally, possibly leading to custom-tailored molecules and unprecedented experimental control of molecular physics. As the prime example to date, dimers bound by repulsive atom-atom interactions have been demonstrated experimentally~\cite{WinklerK441} and confirmed theoretically~\cite{WinklerK441,Nygaard77, ValienteM41, Javanainen81, ValienteMX}.

In the conventional single-channel description of the lattice dimers~\cite{ORS05, WinklerK441, WAN08, VAL08EPL, ValienteM41, ValienteM42, ValienteMX, Javanainen81} the atoms have an interaction between them characterized by a scattering length that diverges at the  resonance. The more nuanced two-channel theory~\cite{GRU07, Nygaard77, NygaardN78} asserts that there are also molecules present as an independent degree of freedom. In this view atom pairs may be converted into molecules, and a resonance occurs when the atom pairs and the molecules have the same energy.

In the present communication we formulate and solve the time-independent Schr\"odinger equation for a lattice dimer within the two-channel model. We find a peculiar qualitative change in the molecular physics compared to the single-channel model: while the latter always presents one bound state for the dimer, in the case of sufficiently weak direct interactions between the atoms the two-channel model may exhibit {two} bound states, one below and one above the band of dissociated dimer states. In particular, while the single-channel description provides for a bound state below the continuum band for attractive atom-atom interactions {\em or\/} a bound state above the dissociation continuum for repulsive interactions, the two bound states for the two-channel dimer may be viewed under certain conditions as the analogs of attractively {\em and\/} repulsively bound pairs occurring simultaneously for the same system parameters.

We begin with the Bose-Hubbard Hamiltonian for a one-dimensional lattice in the single-channel picture~\cite{Javanainen81},
\vspace{-10pt}
\beq
\frac{H_s}{\hbar} \! = \! - \frac{J}{2} \sum_k \left( a^\dagger_{k+1} a_k  \!+ \! a^\dagger_{k-1} a_k \right)  \!+ \! \frac{U}{2} \sum_k a^\dagger_k a^\dagger_k a_k a_k \, .
\label{SMH}
\eeq
The index $k$ runs over the lattice sites, $L$ of them. We use periodic boundary conditions, so that the site $k = L$ is the same as the site $k=0$. The parameter $J$ characterizes tunneling from site to site, $U$ quantifies the atom-atom interactions, and $a_k$ is the annihilation operator for bosonic atoms at the site $k$.

The corresponding two-channel model describes an association resonance where a bound molecular state and a state of two asymptotically free atoms become degenerate. The Hamiltonian is
\bea
\frac{H}{\hbar} & = & - \frac{J}{2} \sum_k \left( a^\dagger_{k+1} a_k + a^\dagger_{k-1} a_k \right) + \delta \sum_k b^\dagger_k b_k \nonumber \\
& & - \xi \sum_k \left( b^\dagger_k a_k a_k + a^\dagger_k a^\dagger_k  b_k \right) \, ,
\label{TMH}
\eea
where $b_k$ is the annihilation operator for a molecule at the site $k$, and the detuning $\delta$ quantifies the difference in energy between a molecule and an atom pair; $\delta = 0$ denotes the resonance. The entities annihilated by $a_k$ and $b_k$ are {bare} atoms and molecules that would exist without the atom-molecule conversion $\propto\xi$. Diagonalization of the Hamiltonian~\eq{TMH} gives a description of the physically observable {dressed} molecules.

In our simplest possible model we do not include site-to-site tunneling of the molecules. This is reasonable since the molecules are twice as heavy as the atoms and the tunneling amplitude $J$ is exponentially small in mass. There could also be atom-atom interactions as in the single-channel model~\eq{SMH}, as well as interactions between molecules and between atoms and molecules, but these are all ignored in the Hamiltonian~\eq{TMH}. In Eq.~\eq{TMH}, and in most of this Rapid Communication, we assume that  the resonance at $\delta \simeq 0$ dominates the physics to the extent that direct two-particle interactions can be ignored.

Although the single-channel model speaks of only atoms that interact among themselves and the two-channel model shows atom-molecule conversion with seemingly no atom-atom interactions at all, there is a close connection between the two descriptions. To demonstrate this, we write the Heisenberg equations of motion for $a_k$ and $b_k$ from the Hamiltonian~\eq{TMH}, and for an asymptotically large detuning $\delta$ eliminate adiabatically the molecular operators $b_k$. With the identification
\beq
U \equiv -{2 \xi^2}/{\delta} \, ,
\label{TCL}
\eeq
the result is the same atomic dynamics as per the  single-channel Hamiltonian~\eq{SMH}. One may think of the single-channel model as the limit of the two-channel model far away from the resonance.

The analysis of the two-channel model proceeds much along the lines of the single-channel case~\cite{Javanainen81}; the multitude of mathematical complications is essentially the same, and we only outline the main points. First, using the discrete Fourier transformation we take both the atomic and the molecular operators into the lattice momentum representation, $a_k\rightarrow c_q$ and $b_k\rightarrow d_q$.
The quasimomentum $q$ runs over a suitable set of values, e.g., $q = 2 \pi n / L$ for integers $n$ such that $q$ belongs to the first Brillouin zone of the lattice. The Hamiltonian becomes
\bea
\frac{H}{\hbar} & = &  - J \sum_q \cos q \, c^\dagger_q c_q + \delta \sum_q d^\dagger_q d_q \nonumber \\ & & - \frac{\xi}{\sqrt{L}} \sum_{q_1,q_2} \left( d^\dagger_{q_1 + q_2} c_{q_1} c_{q_2} + c^\dagger_{q_1} c^\dagger_{q_2} d_{q_1 + q_2} \right) \, .
\eea
For diatomic molecules, the total atom number
\beq
N = \sum_k \left( a^{\dagger}_k a_k + 2 b^{\dagger}_k b_k \right) = \sum_q \left( c^{\dagger}_q c_q + 2 d^{\dagger}_q d_q \right)
%\vspace{-10pt}
\eeq
is a conserved quantity for the two-channel Hamiltonian.

Given two lattice momenta $q_1$ and $q_2$, we next define the analogs of the center-of-mass (c.m.) and relative momenta $P=q_1 + q_2$ and $q = \left( q_1 - q_2 \right) / 2$, and write an ansatz for the state with the total number of atoms $N = 2$ as
\beq
\ket{\psi} = \sum_q A(q) \, c^{\dagger}_{\half P + q} c^{\dagger}_{\half P - q} \ket 0 + \beta \, d^{\dagger}_P \ket 0 \, .
%\vspace{-5pt}
\label{SCHRANSATZ}
\eeq
Here $\ket0$ is the particle vacuum, and without restricting the generality we set $A(q)=A(-q)$. The Hamiltonian maps the state~\eq{SCHRANSATZ} to a state of the same form, with the same $P$, which is a manifestation of conservation of the c.m.\ momentum. The problems with the notation $\half P\pm q$~\cite{Javanainen81} are irrelevant in our final limit when the momentum becomes a continuous variable. Defining a characteristic frequency $\Omega_P$ and scaling all dimensional quantities to it,
\beq
\Omega_P\! =\! 2 J \cos(\half P);\,\,\K = \frac{\sqrt{2}\, \xi}{\Omega_P},\,\,\Delta = \frac{\delta}{\Omega_P},\,\,\omega = \frac{E}{\hbar\Omega_P},
\eeq
we find the time independent Schr\"odinger equation for the energy $E\leftrightarrow\omega$ in the form
\bea
(\omega+\cos q )A(q) +\frac{\K \beta }{\sqrt{2L}} &=& 0,\label{SCHR1}\\
(\Delta-\omega)\beta - \frac{\sqrt{2}\K}{\sqrt{L}}\sum_q A(q) &=& 0\label{SCHR2}\,.
\eea

From the Schr\"odinger equations~\eq{SCHR1} and~\eq{SCHR2} one may deduce the condition for the (scaled) eigenstate energy $\omega$,
%\vspace{-10pt}
\beq
f(\omega, L)\equiv\frac{1}{L}\sum_q \frac{1}{\omega+\cos q}=\frac{\omega - \Delta}{\K^2}\,.
\label{TMEQ}
%\vspace{-10pt}
\eeq
For comparison, the single-channel result is
\beq
f(\omega, L)=1/{\cal K}\,,
\label{SMEQ}
\eeq
with ${\cal K}=U/\Omega_P$.
Once the eigenenergy $\omega$ has been solved from~Eq.~\eq{TMEQ}, the unit-normalized solution to the Schr\"odinger equation is
\beq
\beta = - \left[1 + \frac{\K^2}{L} \sum_q \frac{1}{\left( \omega + \cos q \right)^2} \right]^{-1/2} \, ,
\label{BETAEQ}
\eeq
\vspace{-0.6cm}
\beq
A(q) = \frac{- \K \beta}{\sqrt{2L} \left( \omega + \cos q \right)} \, .
\label{AEQ}
\eeq

\begin{figure}
\includegraphics[width=8.0cm]{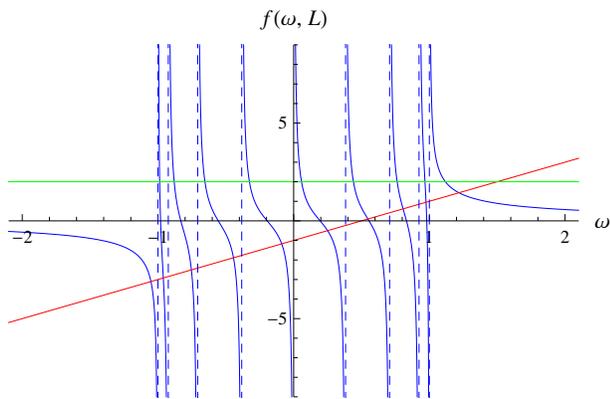}
\caption{(Color online) The function $f(\omega,L)$ [Eq.~\eq{TMEQ}] for $L=16$. The horizontal and positive-slope lines represent the right-hand sides of Eqs.~\eq{SMEQ} and~\eq{TMEQ}, respectively, for $\Kappa = \Delta = \half$ and $\K =1/\sqrt2$. The dashed vertical lines are the asymptotes of $f(\omega,L)$ at the values of $\omega=-\cos{q}$ such that $f(\omega,L)=\pm \infty$.}
\label{FFIG}
%\vspace{-10pt}
\end{figure}

We plot the function $f(\omega, L)$ for $L=16$ as a function of the variable $\omega$ in Fig.~\ref{FFIG}.
The right-hand sides of Eqs.~\eq{TMEQ} and~\eq{SMEQ} for the parameters $\Kappa = \Delta = 1/2$ and $\K = 1/\sqrt2$ are also represented as straight lines in Fig.~\ref{FFIG}. The solutions occur where the straight lines and the graph of $f(\omega, L)$ intersect. Equations~\eq{TMEQ} and~\eq{SMEQ} both have one solution $\omega$ between each value of $-\cos{q}$ for the successive discrete values of the quasimomentum $q$. This is the finite-lattice analog of the dissociation continuum of the dimer. The single-channel eigenvalue equation~\eq{SMEQ} also has one solution $\omega$ outside of the range $(-1,1)$. For a positive atom-atom interaction coefficient $U$ this is the celebrated repulsively bound dimer~\cite{WinklerK441}. On the other hand, the two-channel model  as in Eq.~\eq{TMEQ} always has {\em two} bound states; one above and one below the dissociation continuum.

With the addition of the explicit molecular degree of freedom, for each fixed c.m.\ momentum $P$ the dimer system has one more degree of freedom in the two-channel case than in the single-channel case, which accounts for the existence of the extra state. However, the nature of the newly emerged bound state calls for an investigation.

We discuss the limit when the lattice is infinitely long, $L\rightarrow\infty$, whereupon the quasimomentum becomes a continuous variable in the interval $[-\pi,\pi)$. The main technical rule is that any sum over the lattice momenta is replaced with an integral, $\sum_q \rightarrow \frac{L}{2\pi} \int_{-\pi}^{\pi} dq\,$. Here we only look into the two bound states with $|\omega|>1$.
The eigenvalue equation and the amplitude of the molecules in the bound state are
\beq
\frac{\omega - \Delta} {\K^2 } = \frac{{\rm sgn} \left( \omega \right)}{\sqrt{\omega^2 - 1}} \, ,
\label{CONTBS}
\eeq
\vspace{-0.7cm}
\beq
\beta = - \left[ \frac{\left( \omega^2 - 1 \right)^{3/2}}{\left( \omega^2 - 1 \right)^{3/2} + \abs{\omega} \K^2}  \right]^{1/2} \, .
\eeq

References~\cite{Nygaard77} and~\cite{NygaardN78} based on scattering theory also report multiple bound states, but the explicit results are different from ours as these papers incorporate a significant ``background'' atom-atom interaction $\propto U$ as in Eq.~\eq{SMH}. We may add the background interaction into the present two-channel model, and represent it in terms of the same parameter ${\cal K} =  U/\Omega_P$ that was used in Ref.~\cite{Javanainen81}. For instance, the counterpart of Eq.~\eq{CONTBS} then becomes
\beq
\frac{\omega - \Delta} {{\cal K}(\omega-\Delta)+\K^2 } =  \frac{{\rm sgn} \left( \omega \right)}{\sqrt{\omega^2 - 1}}\,.
\label{FULLEQ}
\eeq
The advantage of the present approach is that, once we have both the bound states and the dissociated states, we can straightforwardly analyze~\cite{Javanainen81} quantities such as the dissociation rate of the bound state when the lattice parameters are modulated, as measured in Ref.~\cite{WinklerK441}. However, the continuum states present similar mathematical issues as in Ref~\cite{Javanainen81}, and we defer the discussion of the dissociation rates for later.

The variety of special cases that may arise with ${\cal K}\ne0$ complicates the discussion, so we continue without the background interactions and set ${\cal K}=0$. Given the actual parameter values, the number of real roots to Eq.~\eq{FULLEQ} with $|\omega|>1$ may be employed as a test of the qualitative validity of the present arguments. For instance, on exact resonance $\Delta=0$ there are two such roots if $\K^2>|{\cal K|}$.

We approximate the bound states as a function of the scaled detuning $\Delta$. First, we note that if the pair $(\omega,\Delta)$ satisfies Eq.~\eq{CONTBS}, then so does  $(-\omega,-\Delta)$. In view of this invariance, it is convenient to assume first that $\Delta\ge0$, and then use the symmetry to deal with the case $\Delta<0$.  In the limit $\Delta\rightarrow\infty$ the angled straight line in Fig.~1 slides down, and it is evident that the two bound-state energies become $\omega\sim\Delta$ and $\omega\sim-1$.
Let us first study the state with $\omega\sim\Delta$. Since an expansion in $\Delta^{-1}$ is expected in the limit $\Delta\rightarrow\infty$, we attempt to find the energy by substituting an expansion of the form $\omega_\Delta = \Delta + a_0 + a_{-1}\Delta^{-1} +\ldots$ into Eq.~\eq{CONTBS}. {\sc mathematica} makes this sort of work extremely easy. The result is
\beq
\omega_\Delta = \Delta + \frac{\K^2}{\Delta}-\frac{\K^2(2\K^2-1)}{2\Delta^3} + {\cal O}\left(\frac{1}{\Delta^{5}}\right)\,.
\label{DD}
\eeq
An expansion of the other bound state energy $\omega_T\simeq-1$ that approaches the continuum threshold in the limit $\Delta\rightarrow\infty$  is found similarly. Analytical expansions are also available in the neighborhood of the association resonance $\Delta\sim0$. In this case the energies $\omega_\pm$ are found in the form $\omega_\pm = \pm\, {\rm sgn}(\omega)\, b_0 + b_1 \Delta + \ldots$.

The extra bound state can be characterized starting from the asymptotic expansions of the bound-state energies. For large detuning the expansions~\eq{DD} and its counterpart for the energy $\omega_T$ give the molecular fractions $f=|\beta|^2$
\beq
f_\Delta \simeq 1 - \frac{\K^2}{\Delta^2}, \quad \,\, f_T \simeq \frac{\K^4}{\abs{\Delta}^3} \, ,
\eeq
whereas for $\Delta\sim0$ the molecular fractions for both bound states are
\beq
f_\pm \simeq \frac{2 \K^4}{1 + 4\K^4 + \sqrt{1 + 4\K^4}} \, .
\label{D0MF}
\eeq

According to Eq.~\eq{TCL}, the single-channel model is recaptured from the two-channel model in the limit when the parameters $\Delta$ and $\K$ both tend to infinity in such a way that ${\cal K} = - \K^2/\Delta$ remains constant. ${\cal K}$ is then nothing but  the two-channel counterpart of the dimensionless atom-atom interaction constant that we also denoted by ${\cal K}$ in our single-channel theory~\cite{Javanainen81}. In fact, in this particular limit the energy of the bound state $\omega_T$ converges to the  single-channel result $\omega={\rm sgn}({\cal K}) \sqrt{1+{\cal K}^2}$. Moreover, the molecular fraction $f_T$ of this bound state vanishes. The $|\Delta|\rightarrow\infty$ bound state that we have denoted by the subscript $T$ is the counterpart in the two-channel model of the bound state in the single-channel model.

The character of the other $|\Delta|\rightarrow\infty$ bound state that we have denoted by the subscript $\Delta$ is equally obvious. The energy tends to $\omega_\Delta\rightarrow\Delta$, the energy of the bare molecule, and the molecular fraction behaves as $f_\Delta\rightarrow1$. This bound state simply represents a bare molecule that has decoupled from the atoms.

The bound states at resonance $\Delta\sim0$ make a more interesting tale. First, the  structure of the state space and of the coupling between bare atoms and molecules, the kinematics of the problem, forces the existence of two bound states for the dressed molecule.  Second, from Eq.~\eq{D0MF} we see that the fraction of bare molecules $f_\pm$ in both bound states tends to zero when the atom-molecule coupling $\propto\K$ vanishes, and to $1/2$ when the atom-molecule coupling is strong. In dimensional quantities the borderline between the two cases is approximately at $\xi\simeq\Omega_P$, when the strength of the atom-molecule coupling is comparable to the width of the continuum band of the dissociated states of the molecule.

In the limit of strong atom-molecule coupling, $|\K|\rightarrow\infty$, the width of the continuum band is negligible. Formally, $\cos{q} \equiv 0 $ in Eq.~\eq{SCHR1} and an effective two-level system for the amplitude $\beta$ and the collective amplitude $\sum_q A_q$ emerges. The association resonance then splits these two states apart. All told, the corresponding bound states represent association in a system that behaves as if there were no kinetic energy for the atoms. On the other hand, for weak atom-molecule coupling, $\K\rightarrow0$, the bound states are already far-detuned from the association resonance as a result of the width of the continuum band, and are effectively described by the single-channel theory. The two coexisting bound states we have denoted by $\pm$ then mean that the usual bound state for an attractive atom-atom interaction and the repulsively bound state are present simultaneously.

The two bound states also provide unexpected insights into modeling of Feshbach resonances. In molecular physics it is customary to think of Feshbach resonances in terms of multiple channels, whereas single-channel pictures are the norm in condensed-matter physics. Oddly enough, there seems to be little difference~\cite{DIE04} between the predictions from single- and two-channel theories in common experimental situations with quantum degenerate gases. In contrast, one bound state in the single-channel description and two bound states in the two-channel description is a stark qualitative difference.

The characteristic frequency scale of the lattice physics~\cite{MOR06} is the recoil frequency set by the atomic mass $m$ and the lattice spacing $d$ as $\epsilon_R = \pi^2 \hbar/2md^2$, roughly $10 \times 2 \pi \, {\rm kHz}$; the  tunneling amplitude $J$ is typically a fraction thereof. The two bound states are similar in character and therefore presumably easiest to detect simultaneously on resonance, $\Delta\sim0$. To have the bound states well separated from the dissociation continuum we would like to have $\xi\gtrsim\Omega_P\sim J$. On the other hand, the resonance will overwhelm the lattice physics if $\xi\gg \epsilon_R$, and the Bose-Hubbard model itself may need to be amended~\cite{DUA05,BUC10}. Overall, it appears that the best experimental parameters are in the neighborhood of $\xi\sim\epsilon_R$. Our question is, are atom-molecule couplings of this order available in practice?

We proceed along the lines of Refs.~\cite{JAV05} and~\cite{MAC10}. We write the atom-molecule coupling as $\xi = \bar\omega^{1/4}\Omega^{3/4}$, where $\bar\omega$ characterizes the free-space molecular physics and $\Omega$ the lattice physics. Assuming that the Wannier functions for atoms and molecules at each lattice site are ground states of a three-dimensional harmonic oscillator with frequencies $\omega_i$, we have $\Omega=(\omega_1\omega_2\omega_3)^{1/3}/2\pi$. The characteristic order of magnitude of the lattice contribution is then of the order of the recoil frequency and, in turn, so should be the free-space contribution, which calls for a weak association resonance.

For the Feshbach resonance, the free-space coupling for a given resonance is fixed once and for all. The remaining molecular frequency is $\bar\omega={4 \pi^2 m a_{\rm b}^2\Delta_B^2\Delta_\mu^2}/{\hbar^3}$, where the background scattering length is $a_{\rm b}$ and the magnetic field width of the resonance is $\Delta_B$, while the difference between the magnetic moments of a bare molecule and two bare atoms is $\Delta_\mu$. To achieve $\bar\omega\sim\epsilon_R$ requires a narrow Feshbach resonance. Potential candidates include the 853~G resonance in Na~\cite{INO98} ($\bar\omega\simeq 1.6 \times 2\pi \, {\rm kHz}$), the 911~G resonance in $^{87}$Rb~\cite{MAR02} ($31\times 2\pi \, {\rm Hz}$), and the 20~G resonance in $^{133}$Cs~\cite{MAR10} $(75 \times 2\pi \, {\rm Hz}$).

In photoassociation the atom-molecule coupling $\xi$ is adjustable according to the laser intensity. Borrowing from Ref.~\cite{MAC10}, we have $\bar\omega={4 \pi^2 m L_{\rm PA}^2\Gamma^2}/{\hbar}$, where $L_{\rm PA}=mK/(4\pi)$ is determined from the low-intensity rate constant $K\propto I$, and the natural linewidth of the molecular state is $\Gamma$. The broad natural linewidths complicate photoassociation in alkali metals~\cite{THE04}, but the alkali-earth metals have narrow linewidths and, in particular, the 192~MHz line in $^{174}$Yb~\cite{YAM10} is a ready candidate for a laser intensity of the order of 1~mW/cm$^2$, so that $\bar\omega\simeq 63 \times 2\pi \, {\rm Hz}$.

We have discussed the stationary states of two bosons in a one-dimensional optical lattice within the two-channel Bose-Hubbard model. The particular result that there are two bound states, one of them effectively a dimer bound by attractive interactions and the other by repulsive interactions, provides a dramatic example of molecular physics in a lattice with no counterpart in free space. We have identified several Feshbach and photoassociation systems for which this unusual situation might be observable.

The authors acknowledge Rekishu Yamazaki for helpful discussions. This work is supported in part by National Science Foundation Grants No.~PHY-0967644 (J.C.S., O.O., and J.J.) and No.~PHY-00900698~(M.M.).

\bibliography{JCSLatticeAnalyticsV130}

\end{document}